\documentclass[a4paper]{article}
\usepackage{amsmath,amssymb}
\usepackage{amscd}
\usepackage{amsfonts}
\usepackage{textcomp}
\usepackage{slashed}
\usepackage{cancel}
\usepackage{pb-diagram}
\usepackage{graphicx}
\usepackage{color}
\usepackage[usenames,dvipsnames,svgnames,table]{xcolor}
\usepackage[bookmarks=false,colorlinks=true,linkcolor={red},citecolor=blue,pdfstartview={XYZ null null 1.22},backref,pagebackref]{hyperref}
\usepackage{cleveref}
\renewcommand{\cal}[1]{\mathcal{#1}}
\begin{document}
\title{Wallis formula from the harmonic oscillator}

\author{Ignacio Cortese and J. Antonio Garc\'ia\\
\small Departamento de F\'isica de Altas Energ\'ias, Instituto de Ciencias Nucleares\\
\small Universidad Nacional Aut\'onoma de M\'exico,\\
\small Apartado Postal 70-543, Ciudad de M\'exico, 04510, M\'exico\\
\small nachoc@nucleares.unam.mx, garcia@nucleares.unam.mx}

\maketitle 

\begin{abstract}
We show that the asymptotic formula for $\pi$, the Wallis formula, that was related with quantum mechanics and the hydrogen atom in \cite{HF}, can also be related to the harmonic oscillator using a quantum duality between these two systems.  As a corollary we show that this very interesting asymptotic formula is not related with the hydrogen atom or quantum mechanics itself but with a clever choice of a trial function and a potential in the Schroedinger equation when we use the variational approach to calculate the ground state energy associated with the given potential function.
\end{abstract}
\maketitle
\section{Introduction}

In the recent article \cite{HF} the authors claim that a venerable asymptotic formula for $\pi$, the Wallis (1655) product, is related to quantum mechanics. More precisely to the variational method applied to the ground state of the hydrogen atom in the  limit of large  angular momentum (the classical limit $\ell\to\infty$). The ingredients used are then a trial function for the variational method, the hamiltonian of the hydrogen atom and the classical limit. A recent analysis \cite{CS} show that if we change the trial function we can still  obtain the Wallis formula or variants of it. The purpose of the present note is to show that the hydrogen atom hamiltonian is also not necessary to obtain the Wallis formula. We will show it explicitly using a well known duality between different potential functions in the Schroedinger equation. So perhaps there are other potentials and trial functions that can give rise to the Wallis asymptotic formula for $\pi$ or variants of it. The classical limit that we will take is just the classical limit used in \cite{HF}.

The Wallis product formula comes from a wise choice of the trial function in the variational approach and the potential function used in the Schroedinger equation. So nothing is magical here. It is interesting to see how some media reacted about this result and the cloud of misunderstanding  that these texts induce \cite{links}. Wallis formula comes from information that we put by hand (the trial function) and nothing reveals that this very nice product formula is inter constructed in quantum mechanics or the hydrogen atom themselves.

\section{Duality in the Schroedinger equation}
Consider the stationary  Schroedinger equation in $d$ spatial dimensions for a  spherically symmetric potential  $V(r)= K\, r^\beta$ with $K$ and $\beta$ some given constants. The radial part is
\begin{equation}\label{S1}
\left(-\frac{\hbar^2}{2m}\left(\frac{d^2}{dr^2}+\frac{(d-1)}{r}\frac{d}{dr}-\frac{\ell(\ell+d-2)}{r^2}\right) +V(r)-E\right)R(r)=0.
\end{equation}
By the change of variable
\begin{equation}
r= \rho^{2/(\beta+2)},
\end{equation}
the Schroedinger equation becomes
\begin{multline}\label{S1bis}
\Bigg(-\frac{\hbar^2}{2m}\left(\frac{d^2}{d\rho^2}+\frac{(\beta+2(d-1))}{(\beta+2)\rho}\frac{d}{d\rho}-\frac{\ell(\ell+d-2)}{\rho^2}\left(\frac{2}{\beta+2}\right)^2\right) \\
-\left(\frac{2}{\beta+2}\right)^2\rho^{\frac{-2\beta}{\beta+2}}E+\left(\frac{2}{\beta+2}\right)^2 K\Bigg)\tilde R(\rho)=0,
\end{multline}
where  $\tilde R(\rho)\equiv R(r(\rho))$.

Now use the following dictionary (duality) to relate the relevant quantities $E,K,d,\ell$ that define the Schroedinger equation to a new set of quantities ${\cal E}, \cal{V},D,{\cal L}$ that parametrize a new system:\\
$\bullet$ dimensions\footnote{The case $d=2$ (conformal point) deserves special attention. See \cite{hoj} for details.}
\begin{equation}\label{D}
D=\frac{2(\beta+d)}{\beta+2}
\end{equation}
$\bullet$ energy
\begin{equation}
{\cal E}= -(\frac{2}{\beta+2})^2 K
\end{equation}
$\bullet$ angular momentum\footnote{Here we are restricting ourselves to the case of integer new dimension $D$ and integer new momenta ${\cal L}$.}
\begin{equation}
{\cal L}= \frac{2}{\beta+2} \ell
\end{equation}
$\bullet$ and the potential function
\begin{equation}\label{V}
{\cal V}= - E \left(\frac{2}{\beta+2}\right)^2\rho^{-\frac{2\beta}{\beta+2}}.
\end{equation}
Using this dictionary, equation (\ref{S1bis}) reads
\begin{equation}
\left(-\frac{\hbar^2}{2m}\left(\frac{d^2}{d\rho^2}+\frac{(D-1)}{\rho}\frac{d}{d\rho}-\frac{\cal{L}(\cal{L}+D-2)}{\rho^2}\right) +{\cal V}(\rho)-{\cal E}\right)\tilde R(\rho)=0,
\end{equation}
which has the \textit{same} form as the Schroedinger equation for potential $\mathcal{V}(\rho)$ in $D$ dimensions, with energy ${\cal E}$ and angular momentum ${\cal L}$. This means that the map $r\to \rho$ and $R(r)\to\tilde{R}(\rho)$, together with (\ref{D}-\ref{V}), produces a solution for the problem with potential $\mathcal{V}(\rho)$ from a solution with potential $V(r)$.

Notice that we cannot map {\em every} solution of the stationary old problem into a stationary solution of the new problem by this duality. We can only map every {\em bounded}  stationary solution of the old problem in a bounded stationary solution of a new problem.

 We will use this dictionary for the case of the hydrogen atom, that is $\beta=-1$, in $d=3$. In such case the new problem is the isotropic harmonic oscillator in $D=4$ and angular momenta ${\cal L}=2\ell$.

\section{Relation with the Wallis formula}

The relation between the hydrogen atom and the Wallis formula is constructed from the trial function\footnote{We will not write the angular part because it will not play any role in the discussion.} 
\begin{equation}\label{trial}
R(r)=e^{-a r^2} r^\ell.
\end{equation}
The analysis given in \cite{HF} evaluates the minimum mean value of the hamiltonian in $d=3$ with the trial function (\ref{trial}) 
\begin{equation}
\langle H\rangle^{\ell}_{min}=-\frac{m e^2}{2\hbar^2}\frac{1}{(\ell+\frac{3}{2})}\Big(\frac{\Gamma(\ell+1)}{\Gamma(\ell+\frac{3}{2})}\Big)^2,
\end{equation}
and from the exact result of the energy eigenvalues takes just the ground state ($N=0$ for $N$ the radial quantum number):
\begin{equation}
E_0^\ell=-\frac{me^4}{2\hbar^2} \frac{1}{(\ell+1)^2}.
\end{equation}
Then the classical limit of the ratio of these two quantities is computed:
\begin{equation}\label{ratio}
\lim_{\ell\to\infty}\frac{\langle H\rangle^{\ell}_{min}}{E^{\ell}_{0}}=\lim_{\ell\to\infty}\frac{(\ell +1)^2}{(\ell +\frac{3}{2})}\Big(\frac{\Gamma(\ell+1)}{\Gamma(\ell+\frac{3}{2})}\Big)^2=1.
\end{equation}
From here the Wallis product formula can be constructed using standard identities for the $\Gamma$ function\footnote{See Appendix B for more details.}
$$\frac{\pi}{2}=\lim_{\ell\to\infty}\prod_{j=1}^{\ell+1}\frac{(2j)(2j)}{(2j-1)(2j+1)}.$$

As stated in \cite{HF} this result is the same for any odd dimensions $d=2k+1$, and the reciprocal of the Wallis formula is obtained for the case of even dimensions $d=2k$. Interestingly enough the relation reported in \cite{HF} for the Wallis formula is symmetric under the interchange between $\ell$ and $k$ so it is possible to interchange the classical limit $\ell\to \infty$ with the limit $k\to \infty$ making contact with quantum mechanics in the large $k$ limit (see for example \cite{LN}). Hence, Wallis formula could also be obtained from large dimensional quantum mechanics. (See the comment at the end of the following section.)

\section{Dual of the Wallis formula}

We start from the hydrogen atom in $d=3$, so $K=-e^2$ and $\beta=-1$. The potential in the Schroedinger equation (\ref{S1}) is $V(r)=-e^2/r$.

Using the duality defined in Sect. 2 we have the relations
\begin{itemize}
\item
$$r=\rho^2$$
\item
$$D=4$$
\item
$${\cal L}=2\ell$$
\item
$$\cal{V}=-4 E \rho^2$$
\item
\begin{equation}\label{newE}
\cal {E}= -4 K= 4 e^2.
\end{equation}
\end{itemize}
The trial function (\ref{trial}) transforms into
\begin{equation}\label{trial2}
\tilde R(\rho)=e^{-a \rho^4} \rho^{\cal L}.
\end{equation}

From the exact result for the energy spectrum of the hydrogen atom
$$ E^\ell_{N}=-\frac{m e^4}{2\hbar^2}\frac{1}{(N+\ell+1)^2}
$$
 we can obtain
 $$
 e^2=\hbar \sqrt{-\frac{ 2 E_{N}}{m} }(N+\ell+1).
 $$
 Then, from (\ref{newE}) the energy of the harmonic oscillator in $D=4$ is written
$$ 
 {\cal E}_n^{\cal L}=\hbar \Bigg(2\sqrt{-\frac{  2 E_{N}}{ m} }\Bigg)(n+\cal{L}+2)=\hbar\omega(n+\cal{L}+2),
 $$
where the definitions $n=2N$ and $\omega=\sqrt{-\frac{  8 E_{N}}{m} }$ have been used.

The exact ground state energy of the harmonic oscillator to consider is then
$$
{\cal E}_0^{\cal L}=\hbar\omega({\cal L}+2).
$$
The variational approximation for the ground state of the harmonic oscillator using the trial function (\ref{trial2}) is (see Appendix A)
\begin{equation}\label{Hmean}
{\langle H\rangle^{{\cal L}}_{min}}=2\hbar\omega\sqrt{\frac{{\cal L} +3}{2}}\Big(\frac{\Gamma(\frac{{\cal L}+3}{2})}{\Gamma(\frac{\cal L}{2}+1)}\Big),
\end{equation}
and the ratio of this variational approximation to the exact ground state energy in the large angular momentum limit gives
\begin{equation}
\lim_{{\cal L}\to\infty}\frac{\langle H\rangle^{{\cal L}}_{min}}{{\cal E}^{{\cal L}}_{0}}=\lim_{{\cal L}\to\infty}\frac{2\sqrt{\frac{{\cal L} +3}{2}}}{({\cal L} +2)}\Big(\frac{\Gamma(\frac{{\cal L}+3}{2})}{\Gamma(\frac{\cal L}{2}+1)}\Big)=1.
\end{equation}
Plugging in that $\cal L=2 \ell $ with $\ell$ an integer we can extract the square root of the reciprocal of Wallis formula from
\begin{equation}\label{main}
\lim_{\ell\to\infty}\frac{\sqrt{\ell+\frac{3}{2}}}{(\ell +1)}\Big(\frac{\Gamma(\ell+\frac{3}{2})}{\Gamma(\ell+1)}\Big)=1.
 \end{equation}
This is our main result.

Through the duality relation we are mapping the hydrogen atom in $d=3$ dimensions to the harmonic oscillator in $D= 4$. 
If we were mapping the hydrogen atom in even dimensions the result would be just the square root of that obtained from the hydrogen atom. In $d=4$ for instance we would get an oscillator in $D=6$ and the square root of the Wallis formula directly. 

In fact, in $d$ dimensions the result is still applicable. The ratio eq. (\ref{ratio}) in $d$ dimensions for the hydrogen atom is
 \begin{equation}
\frac{\langle H\rangle^{d,\ell}_{min}}{E^{d, \ell}_{0}}=\frac{(\ell +\frac{d-1}{2})^2}{(\ell +\frac{d}{2})}\Big(\frac{\Gamma(\ell+\frac{d-1}{2})}{\Gamma(\ell+\frac{d}{2})}\Big)^2.
\end{equation}
 Using our dictionary  with $\beta=-1$ we can write
 \begin{equation}\label{doscl}
\frac{\langle H\rangle^{k,{\ell}}_{min}}{{\cal E}^{k, \ell}_{0}}=\frac{\sqrt{(\ell +k+\frac{1}{2})}}{(\ell +k)}\Big(\frac{\Gamma(\ell+k+\frac{1}{2})}{\Gamma(\ell+k)}\Big).
\end{equation}
where $\ell$ and $k$ are related with the angular momentum ${\cal L}$ and the dimension $D$ of the harmonic oscillator. For example, $D=4$ corresponds to $k=1$ in order to recover our main result (\ref{main}).
In the classical limit $\ell\to\infty$ this equation gives the reciprocal of the square root of the Wallis formula, as expected. Observe that eq. (\ref{doscl}) is symmetric under interchange of  $\ell$ and $k$, so we can also interchange the limit trading the large angular momenta with  large dimension $k\to\infty$.

 We conclude that the relation between the hydrogen atom and the Wallis formula can be traded with a new relation between the harmonic oscillator and a corresponding dual of the Wallis formula. A straightforward calculation reported in Appendix B proves explicitly that our main result eq. (\ref{main}) is in fact the reciprocal square root of the Wallis formula.

 \section{Appendix A}
 
We work out here the calculation of eq. (\ref{Hmean}) for the 4 dimensional isotropic harmonic oscillator using the variational method for the computation of the ground state energy. Let us start from the
 trial function
 $$R(\rho)=e^{-a \rho^4} \rho^{\cal L}.$$
 This unusual trial function comes from the relations $r=\rho^2$ and $\ell={\cal L}/2$ applied to eq. (\ref{trial}). The 
 Schroedinger equation is then
 \begin{equation}
\left(-\frac{\hbar^2}{2m}\left(\frac{d^2}{d\rho^2}+\frac{3}{\rho}\frac{d}{d\rho}-\frac{\cal{L}(\cal{L}+2)}{\rho^2}\right) +K \rho^2-{\cal E}\right) R(\rho)=0,
\end{equation}
where $K=-4E_{N}$. So we take as our starting point the harmonic oscillator with frequency $\omega$ given by 
$$\omega=\sqrt\frac{2 K}{m}=\sqrt{-\frac{  8 E_{N}}{m} }.$$ 
From here it follows that the mean value of the energy of the ground state is
 $$ \langle H\rangle=\frac{\Gamma \left(\frac{{\cal L}+3}{2}\right) \left(2
   \alpha \hbar^2 ({\cal L}+3)+K m\right)}{ \sqrt{2\alpha} m
   \Gamma \left(\frac{\cal L}{2}+1\right)},$$
   where the normalization
   $$\int_0^\infty d\rho\rho^3 e^{-2 a \rho^4} \rho^{2{\cal L}}=2^{-\frac{\cal L}{2}-3} \alpha^{-\frac{\cal L}{2}-1} \Gamma
   \left(\frac{\cal L}{2}+1\right)$$
   has been used.

Minimizing this result with respect to $\alpha$ gives
$$\alpha=\frac{K m}{2 \hbar^2 ({\cal L}+3)},$$
and then
$$\langle H\rangle_{min}=2\hbar\omega\frac{  \Gamma
   \left(\frac{{\cal L}+3}{2}\right)}{\Gamma
   \left(\frac{\cal L}{2}+1\right)} \sqrt{ \frac{({\cal L}+3)}{2}}.
$$
This relation is the same as eq. (\ref{Hmean}) in the main text.

\section{Appendix B}

Here we will use elementary $\Gamma$ function identities to show that the Wallis product follows from the limit (\ref{ratio}). The basic $\Gamma$ identities are $n\Gamma(n)=\Gamma(n+1)$, $\Gamma(\frac12)=\sqrt\pi$, $\Gamma(n)=(n+1)!$ and the duplication formula \cite{M2}
$$\Gamma(2n)=\frac{2^{2n-1}\Gamma(n)\Gamma(n+\frac12)}{\Gamma(\frac12)}.$$
Start from the limit
$$
\lim_{\ell\to\infty}\frac{(\ell +1)^2}{(\ell +\frac{3}{2})}\Big(\frac{\Gamma(\ell+1)}{\Gamma(\ell+\frac{3}{2})}\Big)^2=1.
$$
Just define $n=\ell+1$ to get
$$
\lim_{n\to\infty}\frac{n^2}{(n +\frac{1}{2})}\Big(\frac{\Gamma(n)}{\Gamma(n+\frac{1}{2})}\Big)^2=1.
$$
Observe that $n^2\Gamma(n)^2=\Gamma(n+1)^2$ and $(n+\frac12)=\Gamma(n+\frac32)/\Gamma(n+\frac12)$ so we get
\begin{equation}\label{B1}
\lim_{n\to\infty}\frac{\Gamma(n+1)^2}{\Gamma(n+\frac32)\Gamma(n+\frac12)}=1.
\end{equation}
Now using the duplication formula we have \cite{etopi}
$$\frac{\Gamma(n+1)}{\Gamma(n+\frac12)}= \frac{1}{\sqrt\pi}\frac{(2^n n \Gamma(n))^2}{2n\Gamma(2n)}=\frac{1}{\sqrt\pi}\prod_{j=1}^n\frac{2j}{2j-1}.$$
In the same fashion we obtain
$$\frac{\Gamma(n+1)}{\Gamma(n+\frac32)}= \frac{2}{\sqrt\pi}\frac{(2^n n \Gamma(n))^2}{\Gamma(2n+2)}=\frac{2}{\sqrt\pi}\prod_{j=1}^n\frac{2j}{2j+1}.$$
The last step comes from multiplying the last two results to get
$$\frac{\Gamma(n+1)^2}{\Gamma(n+\frac12)\Gamma(n+\frac32)}=\frac2\pi\prod_{j=1}^n\frac{(2j)(2j)}{(2j+1)(2j-1)},$$
and taking the limit we have the Wallis formula
$$\lim_{n\to\infty}\prod_{j=1}^n\frac{(2j)(2j)}{(2j+1)(2j-1)}=\frac\pi2.$$

By the same token we can show that our main result eq. (\ref{main}) can be related to the reciprocal square root of the Wallis formula. Starting from eq. (\ref{main})
$$
\lim_{n\to\infty}\frac{\sqrt{n+\frac{3}{2}}}{(n +1)}\Big(\frac{\Gamma(n+\frac{3}{2})}{\Gamma(n+1)}\Big)=1,
$$
with the definition $m=n+1$ we have
$$
\lim_{m\to\infty}\frac{\sqrt{m+\frac{1}{2}}}{m}\frac{\Gamma(m+\frac{1}{2})}{\Gamma(m)}=1.
$$
Using our previous identities this equation can be written  in the form
$$
\lim_{m\to\infty}\sqrt{\Bigg(\frac{\Gamma(m+\frac32)}{\Gamma(m+1)}\Bigg)\Bigg(\frac{\Gamma(m+\frac12)}{\Gamma(m+1)}\Bigg)}=1,
$$
which has the same form as the previous result eq. (\ref{B1}) but for the reciprocal square root of $\pi/2$.  

\section{Acknowledgements}
The authors were  partially supported by Mexico National Council of Science and Technology (CONACyT) grant 238734 and DGAPA-UNAM grant IN107115.\\

\paragraph{NOTE ADDED} After the first submission of this note, the very nice and detailed exposition in \cite{Paddy} about the duality between the isotropic harmonic oscillator in $D=4$ and the hydrogen atom in $d=3$ was pointed to our attention. Many details of this interesting duality, at classical and quantum level, not covered here are presented with clear and deep insights in that reference.

\end{document}